\documentclass[trackchanges,twocolumn]{aastex701}
\usepackage{CJK}

\newcommand{\cago}{^{12}{\rm C}(\alpha,\gamma)^{16}{\rm O}}
\newcommand{\msun}{{\rm M}_\odot}
\newcommand{\rsun}{{\rm R}_\odot}
\newcommand{\mzms}{m_{\rm zams}}
\newcommand{\mrem}{m_{\rm rem}}
\newcommand{\rate}{{\rm yr}^{-1}\;{\rm Gpc}^{-3}}
\newcommand{\xeff}{\chi_{\rm eff}}
\newcommand{\xpre}{\chi_{\rm p}}
\newcommand{\fdir}{./figure}

\begin{document}
\begin{CJK*}{UTF8}{gbsn}

%%\title{Template \aastex v7.0.1 Article with Examples\footnote{Footnotes can be added to titles}}
\title{GW231123 Formation from Population III Stars: Isolated Binary
  Evolution}

\author[orcid=0000-0002-8461-5517,sname='Tanikawa']{Ataru Tanikawa}
\affiliation{Center for Information Science, Fukui Prefectural University}
\email[show]{tanik@g.fpu.ac.jp}

\author[orcid=0000-0002-1197-2054, sname='Liu']{Shuai Liu (刘帅)}
\affiliation{School of Electronic and Electrical Engineering, Zhaoqing University}
\email{}

\author{WeiWei Wu (吴维为)}
\affiliation{School of Physics and Astronomy, Sun Yat-sen University}
\email{}

\author[orcid=0000-0002-6465-2978,sname='Fujii']{Michiko S. Fujii} 
\affiliation{Department of Astronomy, Graduate School of Science, The University of Tokyo}
\email{fujii@astron.s.u-tokyo.ac.jp}

\author[orcid=0000-0001-8713-0366,sname='Wang']{Long Wang (王龙)}
\altaffiliation{CSST Science Center for the Guangdong-Hong Kong-Macau Greater Bay Area}
\affiliation{School of Physics and Astronomy, Sun Yat-sen University}
\email{wanglong8@sysu.edu.cn}

%%\collaboration{all}{The Terra Mater collaboration}

\begin{abstract}

  GW231123 is a merger of two black holes (BHs) with estimated masses
  exceeding $100\;{\rm M}_{\odot}$, making them the most massive BHs
  discovered to date via gravitational wave (GW) observations.  We
  investigate whether GW231123-like events can originate from isolated
  Population (Pop) III binary stars using binary population synthesis
  calculations.  Our findings indicate that isolated Pop III binaries
  can produce GW231123-like events at a rate sufficient to explain the
  discovery of GW231123, provided that three conditions are met: (i)
  Pop III stars evolve with inefficient convective overshooting, (ii)
  the $^{12}\text{C}(\alpha, \gamma)^{16}\text{O}$ rate is $2\sigma$
  lower than the standard value, and (iii) Pop III binary stars share
  the same orbital parameters as Pop I/II binary stars at the initial
  time. In contrast, GW190521 -- the most massive BH merger in the
  Gravitational Wave Transient Catalog 3 -- can be formed from
  isolated Pop III binaries even with the standard
  $^{12}\text{C}(\alpha, \gamma)^{16}\text{O}$ rate. We demonstrate
  that the discovery of GW231123 is increasingly constraining the
  parameter ranges of single star evolution models, under the
  assumption that these GW events originate from isolated binary
  evolution.

\end{abstract}

\keywords{\uat{Gravitational wave sources}{677} --- \uat{Stellar mass
    black holes}{1611} --- \uat{Astrophysical black holes}{98} ---
  \uat{Population III}{1285} --- \uat{Close binary stars}{254}}


\section{Introduction}
\label{sec:Introduction}

Gravitational wave (GW) observations have uncovered a vast population
of binary black hole (BH) mergers, with nearly 200 events detected as
of the first half of the fourth observing run
\citep[O4a;][]{2025arXiv250818082T}. Despite this growing sample,
their astrophysical origins remain a subject of active debate
\citep{2025arXiv250818083T}. A recently reported event, GW231123
\citep{2025arXiv250708219T}, stands out due to two remarkable
features.  First, it involves the most massive BHs discovered to date,
with inferred masses of $137_{-17}^{+22}\;\msun$ and
$103_{-51}^{+20}\;\msun$.  These masses either fall within or exceed
the pair instability (PI) mass gap of $\sim 60$ -- $130\;\msun$
\citep{2020ApJ...902L..36F, 2021ApJ...912L..31W, 2022ApJ...924...39M,
  2022ApJ...937..112F}, though the secondary BH mass may lie below the
gap depending on the choice of analysis pipeline. Second, both BHs
exhibit potentially high spins, estimated at $0.90_{-0.19}^{+0.10}$
and $0.80_{-0.51}^{+0.20}$ for the primary and secondary components,
respectively. These characteristics suggest several formation
channels, such as hierarchical mergers in dense star clusters
\citep{2019PhRvD.100d3027R, 2021MNRAS.501.5257R, 2021MNRAS.507.5132D,
  2025arXiv250715967S} or within active galactic nucleus (AGN) disks
\citep{2020ApJ...898...25T, 2025arXiv250717551L,
  2025arXiv250908298L}. Alternatively, some studies have proposed that
GW231123 could originate from isolated binary evolution
\citep{2025arXiv250810088C, 2025ApJ...993L..54G, 2025arXiv250900154P,
  2025arXiv250904574B}. It has also been suggested that the GW signals
could be produced by the collapse of a rapidly spinning massive star
in the mass range of $10^2$ -- $10^4\;\msun$
\citep{2021PhRvD.103f3037S, 2025arXiv250915619S}.

Population (Pop) III stars, or metal-free stars, represent a
compelling potential origin for exceptionally massive BH mergers such
as GW231123.  Unlike Pop I and II stars, Pop III stars are thought to
be typically massive, ranging from $10$ to $10^3\;\msun$
\citep{1998ApJ...508..141O, 2002Sci...295...93A, 2004ARA&A..42...79B,
  2008Sci...321..669Y, 2011Sci...334.1250H, 2013ApJ...773..185S,
  2014ApJ...792...32S, 2014ApJ...781...60H, 2015MNRAS.448..568H,
  2016ApJ...824..119H, 2020MNRAS.497..336S, 2021MNRAS.503.2014S,
  2022MNRAS.512..116J, 2023ARA&A..61...65K}, although some
lower-resolution simulations may overestimate these stellar masses
\citep{2022MNRAS.510.4019P}. Consequently, they are expected to leave
behind more massive BHs. Numerical simulations further indicate that a
significant fraction of Pop III stars form in multiple systems,
including binaries and small clusters \citep{2010MNRAS.403...45S,
  2011ApJ...727..110C, 2012MNRAS.424..399G, 2013ApJ...768..131V,
  2015MNRAS.448.1405M, 2015MNRAS.452.1233H, 2017MNRAS.470..898H,
  2019ApJ...877...99S, 2020ApJ...892L..14S}.  These Pop III systems
may contribute to the observed population of binary BH mergers through
either isolated binary evolution \citep{2004ApJ...608L..45B,
  2014MNRAS.442.2963K, 2021ApJ...910...30T, 2023MNRAS.524..307S,
  2023MNRAS.525.2891C} or dynamical interactions within Pop III
clusters \citep{2020MNRAS.495.2475L, 2022MNRAS.515.5106W,
  2024MNRAS.533.2262L, 2024A&A...690A.106M,
  2025ApJ...986..163W}. Notably, it has been suggested that even
GW190521, an event within the PI mass gap, could have originated in
such Pop III environments \citep{2020ApJ...903L..40L,
  2021MNRAS.501L..49K, 2021MNRAS.505.2170T}.

In this paper, we investigate the formation of GW231123-like events
within the framework of isolated Pop III binary evolution. We employ
two distinct stellar evolution models for Pop III stars and explore
various criteria for pair instability supernova (PISN) and pulsational
PISN (PPISN) models. By surveying a wide range of parameters, we aim
to identify the conditions necessary to produce GW231123-like events,
while also ensuring consistency with the broader population of
observed binary BH mergers. In a companion paper
\citep{2025ApJ...993L..30L}, we extend this analysis to the dynamical
formation of GW231123-like events within Pop III star clusters.

Before concluding this section, we define ``GW231123-like events'' as
BH mergers with component masses falling within the 90 \% credible
intervals of GW231123, as derived from the combined model of the five
pipelines adopted in \cite{2025arXiv250708219T}. In other words,
$137_{-17}^{+22}\;\msun$ for the primary BH mass,
$103_{-51}^{+20}\;\msun$ for the secondary BH mass, and
$0.75_{-0.39}^{+0.22}$ for the mass ratio. While our primary
definition does not include effective inspiral spins ($\xeff$), we
nonetheless evaluate the fraction of our synthesized GW231123-like
events that also fall within the 90 \% credible intervals for $\xeff$
reported in the same combined model ($\xeff =
0.31_{-0.39}^{+0.24}$). We do not consider effective precessing spins
($\xpre$), as their determination requires higher-quality data than
those for $\xeff$ and mass ratios; indeed, \cite{2025arXiv250708219T}
note that they cannot confidently claim evidence of precession. The
definition of ``GW190521-like events'' follows a similar methodology:
we identify them as mergers with component masses within the 90 \%
credible intervals of GW190521 reported in \cite{2020PhRvL.125j1102A},
or $85_{-14}^{+21}\;\msun$ for the primary BH mass,
$66_{-18}^{+17}\;\msun$ for the secondary BH mass, and
$0.79_{-0.29}^{+0.19}$ for the mass ratio. We do not adopt additional
constraints on their effective inspiral or precessing spins.

The remainder of this paper is structured as follows. In Section
\ref{sec:Method}, we describe our methodology for tracking binary BH
formation through isolated binary evolution. In Section
\ref{sec:Results}, we present our numerical results and provide a
detailed analysis of the findings. Finally, in Section
\ref{sec:Conclusion_and_Discussion}, we summarize our conclusions and
discuss the implications for the formation of GW231123.

\section{Method}
\label{sec:Method}

We perform binary population synthesis calculations to investigate
binary BH formation through isolated binary evolution. For this
purpose, we employ the BSEEMP code \citep{2020MNRAS.495.4170T,
  2022ApJ...926...83T}, which is a derivation of the original BSE code
\citep{2000MNRAS.315..543H, 2002MNRAS.329..897H}. While the original
BSE code is restricted to metallicities $Z \ge 10^{-4}$, the BSEEMP
code extends this range down to $Z = 2 \times 10^{-10}$, a regime
where stellar evolution is effectively identical to that of Pop III or
metal-free stars. To model the Pop III star formation rate density
(SFRD), we adopt the prescription from Eq. (5) in
\cite{2022ApJ...926...83T}, which is based on the cosmological
simulation results of \cite{2020MNRAS.492.4386S}. The SFRD is
expressed as:
\begin{eqnarray}
  &\log ( \phi_{\rm III} (t) / [\msun \; {\rm yr}^{-1} \; {\rm
        Mpc}^{-3}] ) = \nonumber \\ &\left\{
  \begin{array}{ll}
    0.0130   (t/{\rm Myr} - 100) - 5.00 & (100 \le t/{\rm Myr} < 200) \\
    -0.00261 (t/{\rm Myr} - 200) - 3.70 & (200 \le t/{\rm Myr} < 400) \\
    -0.0108  (t/{\rm Myr} - 400) - 4.22 & (400 \le t/{\rm Myr} < 500) \\
    0                                 & (\mbox{otherwise})
  \end{array}
  \right.. \label{eq:PopIIISfr}
\end{eqnarray}
The resulting total integrated mass density of Pop III stars is $\sim
3 \times 10^{13} \, \msun \, {\rm Gpc}^{-3}$, which does not exceed
the upper limit of Pop III star density constrained by reionization
\citep{2015MNRAS.453.4456V, 2025MNRAS.541.3113L} and GW background
\citep{2016MNRAS.461.2722I, 2021ApJ...919...41I}. Additionally, it is
also consistent with recent observational constraints on the early
star formation rate \citep[e.g.,][]{2025ApJ...980..138H}.

The initial conditions for Pop III binaries are as follows. For the
primary stars, we adopt a top-heavy initial mass function (IMF)
following a power-law index of $-1$ within the mass range of $10$ to
$150\;\msun$. The binary fraction is assumed to be $0.5$, which is
broadly consistent with the observations of massive stars by
\cite{2012Sci...337..444S}. We adopt the distributions derived by
\cite{2012Sci...337..444S} for the distributions for binary mass
ratios, orbital periods, and orbital eccentricities. To ensure
statistical significance, we follow the evolution of 3 million binary
systems. We simulate not only Pop III binary stars but also Pop I and
II binary stars, and thus their initial conditions are detailed in
Appendix \ref{sec:popIandII}.

Our binary evolution models follow the prescriptions of
\cite{2022ApJ...926...83T}, with the exception of the PISN and PPISN
models. Stellar winds are implemented as in
\cite{2010ApJ...714.1217B}, where hot, massive hydrogen-rich stars
undergo mass loss proportional to $Z^{0.85}$ \citep{Vink01}. As noted
above, we treat $Z = 2 \times 10^{-10}$ as the metallicity for Pop III
stars; thus, their stellar wind mass loss is small but
non-zero. Furthermore, Pop III stars can experience stellar wind mass
loss similar to luminous blue variables (LBVs) at a rate of $1.5
\times 10^{-4} \; \msun \; {\rm yr}^{-1}$ if they satisfy the criteria
described in \cite{1994PASP..106.1025H} and
\cite{2010ApJ...714.1217B}. The BH spins are determined based on the
angular momenta of their progenitors and the loss of angular momenta
during gravitational collapse (see Appendix \ref{sec:spin_evolution}
for details). The binary evolution model accounts for tidal evolution,
stable mass transfer, common envelope (CE) evolution, and orbital
decay via GW radiation. We adopt the $\alpha$ formalism for CE
evolution \citep{1984ApJ...277..355W}, where $\alpha = 1$, and the
binding energy parameter $\lambda$ is the same as the model of
\cite{2014A&A...563A..83C}. Stable mass transfer is assumed to be
Eddington-limited. In the sub-Eddington regime, we assume that half of
the transferred mass is lost from the binary system.

Next, we detail the modifications made to the prescriptions of
\cite{2022ApJ...926...83T}, specifically regarding our PPISN, PISN,
and core-collapse supernova (CCSN) models. For CCSNe, we adopt the
``rapid'' supernova engine \citep{2012ApJ...749...91F}. PPISN and PISN
are triggered when the helium core mass ($m_{\rm c}$) of a progenitor
satisfies specific criteria, resulting in a remnant mass ($\mrem$)
defined as follows:
\begin{equation}
  \mrem = \left\{
  \begin{array}{ll}
    m_{\rm rapid} & (m_{\rm c} \le m_{\rm c,PPISN})\\
    m_{\rm c,PPISN} & (m_{\rm c,PPISN} \le m_{\rm c} \le m_{\rm c,PISN})\\
    0 & (m_{\rm c,PISN} \le m_{\rm c} \le m_{\rm c,DC})\\
    m_{\rm rapid} & (m_{\rm c} \ge m_{\rm c,DC})
  \end{array}
  \right.,
\end{equation}
where $m_{\rm c,PPISN}$, $m_{\rm c,PISN}$, and $m_{\rm c,DC}$
represent the lower mass limits for the onset of PPISN, PISN, and
direct collapse (DC), respectively. These thresholds are highly
sensitive to the $\cago$ reaction rate \citep{2018ApJ...863..153T,
  2020ApJ...902L..36F, 2021MNRAS.501.4514C, 2024MNRAS.531.2786K}.
Following the results of \cite{2020ApJ...902L..36F}, we set $(m_{\rm
  c,PPISN}, m_{\rm c,PISN}, m_{\rm c,DC}) = (45, 65, 135)\,M_{\odot}$
for the standard $\cago$ rate. To investigate the impact of nuclear
reaction uncertainties, we also consider rates that are $1\sigma$,
$2\sigma$, and $3\sigma$ lower than the standard value, corresponding
to thresholds of $(55, 75, 145)$, $(85, 85, 160)$, and $(90, 90,
180)\;\msun$, respectively. For simplicity, we assume a single fixed
value for the remnant mass of PPISN, although we acknowledge that the
actual mass loss likely depends on the progenitor's structure in a
more complex manner \citep{2019ApJ...882..121S,
  2022RNAAS...6...25R}. Furthermore, while rotation has been suggested
to influence the location of the PISN mass gap
\citep{2020A&A...640L..18M, 2025arXiv250810088C, 2025arXiv250815887G},
we do not explore its effects in the present study.

For cases where the $\cago$ rate is $2\sigma$ or $3\sigma$ lower than
the standard value, we set $m_{\rm c,PPISN} = m_{\rm c,PISN}$. This
assumption effectively implies that PPISNe do not occur under these
conditions, consistent with the findings of \cite{2020ApJ...902L..36F}
and \cite{2022ApJ...924...39M}. Specifically,
\cite{2022ApJ...937..112F} explicitly stated that PPISNe are
suppressed when the $\cago$ rate is $3\sigma$ lower than the standard
value. While they did not explicitly address the occurrence of PPISNe
for the $2\sigma$ lower rate, we infer a similar absence based on the
fact that the results of \cite{2022ApJ...937..112F} align closely with
those of \cite{2020ApJ...902L..36F} and \cite{2022ApJ...924...39M}.

Our single-star evolution model for $Z < 10^{-4}$ is constructed using
fitting formulae based on 1D hydrodynamics simulations performed with
the HOSHI code \citep{2016MNRAS.456.1320T, 2018ApJ...857..111T,
  Takahashi19, Yoshida19}. The HOSHI code provides two distinct sets
of models characterized by different convective overshooting
parameters: the L and M models, which correspond to efficient and
inefficient overshooting, respectively. The evolutionary tracks of Pop
III stars differ significantly between these two models. For instance,
the maximum radius reached by an $80\,\msun$ star is $\lesssim
10^2\,\rsun$ in the M model, whereas it expands to $\gtrsim
10^3\,\rsun$ in the L model. Further details regarding the HOSHI code
and the specific properties of the M and L models are provided in
Appendix \ref{sec:single_star}.

\begin{deluxetable*}{cccccl}
\tablewidth{0pt}
\tablecaption{Parameter sets and local merger rate densities for
  GW231123- and GW190521-like events. \label{tab:Parameter_sets}}
\tablehead{
  \colhead{Name} & \colhead{Single star model} & \colhead{PPISN/PISN
    model} & GW231123-like event & GW190521-like event \\
  & (Overshoot efficiency) & ($\cago$ rate) & [$\rate$] & [$\rate$]
}
\startdata
Mstd       & M (inefficient) & Standard        & $0$ & $4.9 \times 10^{-2}$ \\
M$1\sigma$ & M (inefficient) & $1\sigma$ lower & $0$ & $1.0 \times 10^{-1}$ \\
M$2\sigma$ & M (inefficient) & $2\sigma$ lower & $2.1 \times 10^{-2}$ & $1.0 \times 10^{-1}$ \\
M$3\sigma$ & M (inefficient) & $3\sigma$ lower & $2.1 \times 10^{-2}$ & $1.0 \times 10^{-1}$ \\
Lstd       & L (efficient)   & Standard        & $0$ & $0$ \\
L$3\sigma$ & L (efficient)   & $3\sigma$ lower & $0$ & $8.6 \times 10^{-2}$ \\
\enddata
\tablecomments{The Mstd and L$3\sigma$ sets correspond to the fiducial
  and L-$3\sigma$ sets in \cite{2022ApJ...926...83T},
  respectively. All merger rate densities are evaluated at redshift $z
  = 0$.}
\end{deluxetable*}

\begin{figure}[ht!]
  \plotone{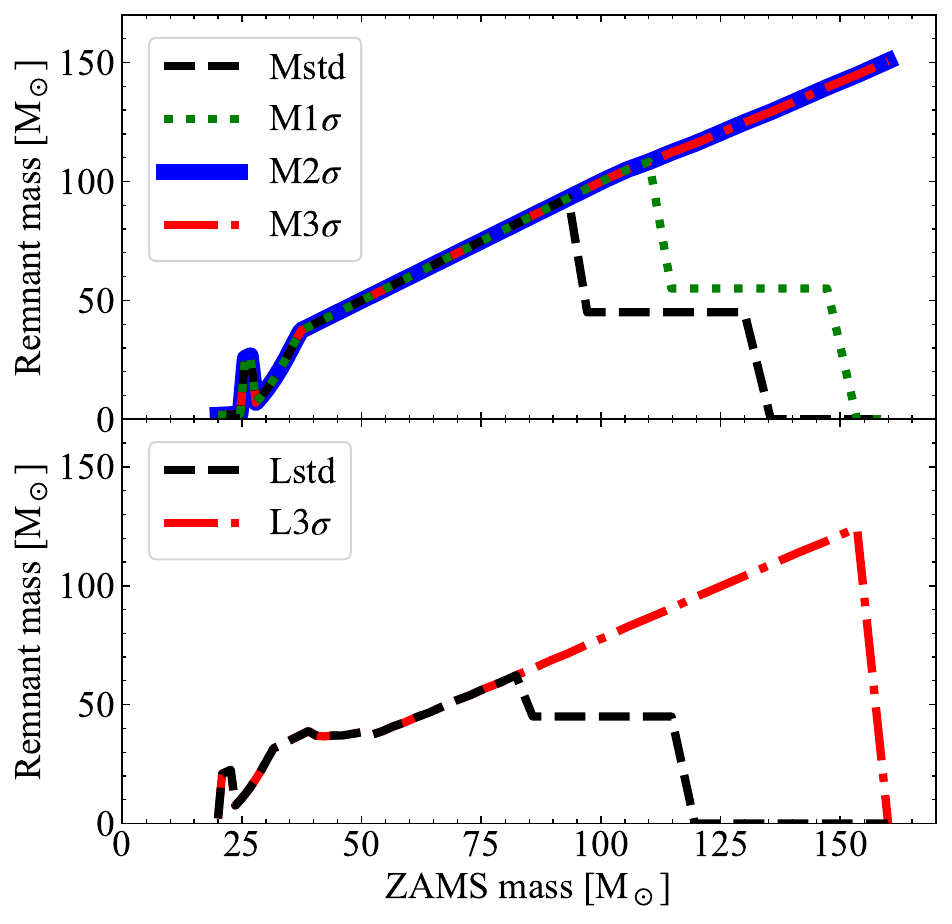}
  \caption{Relation between zero-age main sequence (ZAMS) and
    remnant masses for parameter sets we adopt. The curves of the
    M$2\sigma$ and M$3\sigma$ sets overlap. \label{fig:zamsrem}}
\end{figure}

We investigate 6 parameter sets by combining different single star
evolution models with various PPISN/PISN prescriptions, as summarized
in Table \ref{tab:Parameter_sets}. The model names are assigned based
on these choices: the first letter indicates the stellar evolution
model (M or L), while the subsequent part represents the adopted
$\cago$ rate -- namely, the standard rate (std) or rates that are
$1\sigma$, $2\sigma$, or $3\sigma$ lower. For example, the ``Mstd''
model denotes the combination of the M model for single star evolution
and the standard $\cago$ rate.

The Mstd and L$3\sigma$ sets were previously investigated by
\cite{2022ApJ...926...83T}, who demonstrated that GW190521-like events
-- which fall within the PI mass gap -- can be formed under these
conditions. In this study, we highlight the inherent difficulty of
forming both GW190521 and GW231123 within the framework of the L
model. This difficulty motivated our decision to limit the L model
analysis to only two cases: the standard and $3\sigma$ lower $\cago$
rates. As discussed in Section \ref{sec:Results}, since neither the
Lstd nor the L$3\sigma$ sets produce GW231123-like events, it follows
that the intermediate L$1\sigma$ and L$2\sigma$ cases would also fail
to yield such mergers.

Figure \ref{fig:zamsrem} illustrates the relationship between the
zero-age main-sequence (ZAMS) mass ($\mzms$) and the resulting remnant
mass for the 6 parameter sets. In the Mstd set, the remnant mass is
reduced to $\sim 40\,\msun$ at $\mzms \sim 90\,\msun$ due to PPISN,
and drops to zero at $\mzms \sim 130\,\msun$ due to PISN. As the
$\cago$ rate decreases, the thresholds for PPISN and PISN shift toward
higher masses. Notably, in the M2$\sigma$ and M3$\sigma$ sets, PISN
does not occur for progenitors with $\mzms$ up to $150\,\msun$ (the
upper limit of our IMF). As previously noted, we assume that PPISN is
entirely suppressed for these lower reaction rates. While the Lstd set
follows a similar trend to the Mstd set, the thresholds are shifted
toward lower $\mzms$. This is because the more efficient convective
overshooting in the L model leads to larger helium core masses for a
given $\mzms$ compared to the M model. For the same reason, PISN
occurs at $M_{\rm ZAMS} = 150\,\msun$ even in the L3$\sigma$
set. Throughout our models, Pop III stars experience minimal mass loss
via stellar winds (at most a few percent of the initial mass),
consistent with the findings of \cite{2021MNRAS.504..146V} and
\cite{2022ARA&A..60..203V}.\footnote{Stars in the L and M models
expand to $\gtrsim 10^3\,\rsun$ and $\lesssim 10^2\,\rsun$,
respectively. LBV-like winds are triggered only for stars in the L
model; consequently, these stars leave behind less massive BHs than
those in the M model. Nevertheless, stars in the L models lose their
masses via stellar winds by a few percent of their initial masses.}

\section{Results}
\label{sec:Results}

We investigated 6 parameter sets and summarized the resulting merger
rate densities for GW190521- and GW231123-like events in Table
\ref{tab:Parameter_sets}. In the following sections, we analyze the
M2$\sigma$ and L3$\sigma$ sets in detail as representative cases. It
is important to note that GW231123-like events become increasingly
easier to form as the $\cago$ rate decreases; this is because the
lower $\mzms$ threshold for PISNe shifts toward higher masses and the
$\mzms$ range for PPISNe eventually disappears (see Figure
\ref{fig:zamsrem}). The M2$\sigma$ set is the model with the highest
$\cago$ rate among the M-series models capable of producing
GW231123-like events. Consequently, the M3$\sigma$ set also naturally
produces such events. In fact, the relationships between the ZAMS and
remnant masses are the same between the M$2\sigma$ and M$3\sigma$ sets
in the ZAMS mass range up to $160;\msun$ (see Figure
\ref{fig:zamsrem}. These sets yield the same results, as shown in the
GW231123-like and GW190521-like events in Table
\ref{tab:Parameter_sets}.  In contrast, despite having the lowest
$\cago$ rate, the L3$\sigma$ set fails to produce any GW231123-like
events. This indicates that the other models in the L-series, which
utilize higher $\cago$ rates, are similarly unable to form such
massive mergers.

\begin{figure}[ht!]
  \plotone{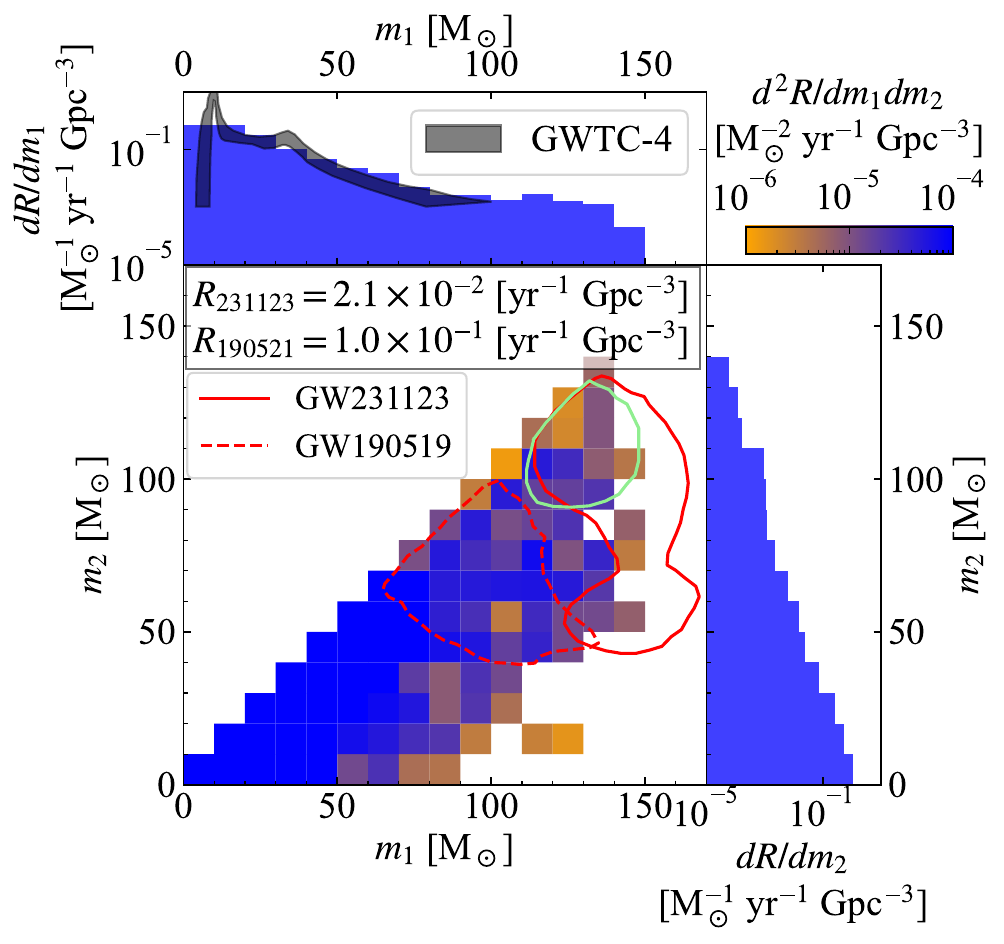}
  \plotone{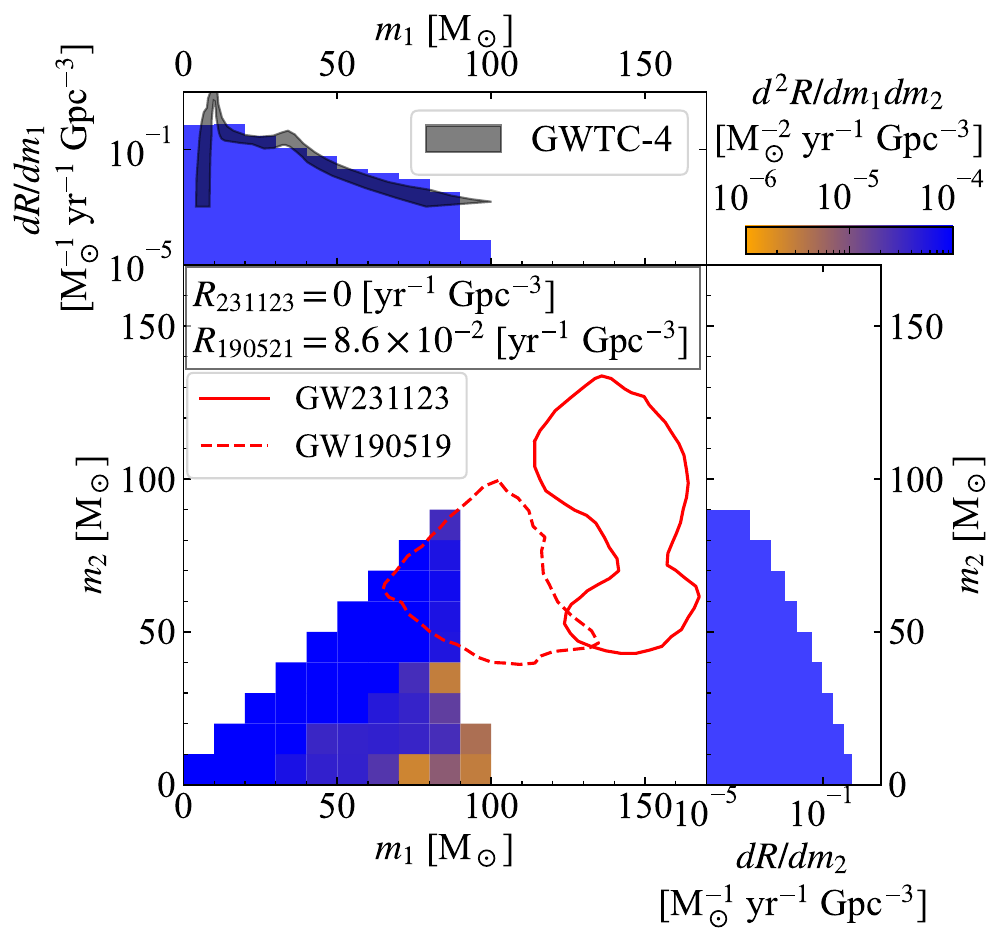}
  \caption{Local merger rate density distribution ($z=0$) as a
    function of primary ($m_1$) and secondary ($m_2$) BH masses for
    the M$2\sigma$ (top) and L$3\sigma$ (bottom) sets. The color scale
    for each panel is indicated by the color bar in the upper
    right. Solid and dashed curves represent the 90 \% credible
    intervals for the BH masses of GW231123 and GW190521, respectively
    \citep{2025arXiv250708219T}. The credible interval for GW231123 is
    derived from the combined model of the five pipelines used in the
    reference study. To estimate $R_{\rm GW231123}$ and $R_{\rm
      GW190521}$, we calculate the total merger rate density
    integrated within these curves, without applying weights based on
    the event posterior distributions
    \citep[e.g.][]{2023MNRAS.525.3986M}. The green curve denotes the
    90 \% credible interval for GW231123 based on the NRSur7DQ4
    waveform \citep{2019PhRvR...1c3015V}, within which the merger rate
    density is approximately $0.017\;{\rm Gpc}^{-3}\;{\rm yr}^{-1}$.
\label{fig:gw231123_rate}}
\end{figure}

Figure \ref{fig:gw231123_rate} displays the binary BH merger rate
density as a function of the primary ($m_1$) and secondary ($m_2$) BH
masses for the M2$\sigma$ and L3$\sigma$ sets. For both models, the
primary mass distributions are consistent with the results from the
Gravitational Wave Transient Catalog 3 (GWTC-3) for $m_1 <
100\,\msun$. We note that the merger rate density for the $m_1 >
100\,\msun$ regime was not explicitly provided in the latest analysis
by \cite{2025arXiv250818083T}. In the M2$\sigma$ set, GW231123-like
events occur at a rate of $2.1 \times 10^{-2}\,{\rm Gpc}^{-3}\,{\rm
  yr}^{-1}$, which is consistent with the single-event rate estimate
of $0.08_{-0.07}^{+0.19}\,{\rm Gpc}^{-3}\,{\rm yr}^{-1}$ reported by
\cite{2025arXiv250708219T}. This consistency holds even if we adopt
the 90 \% credible intervals for BH masses derived from a single
pipeline using the NRSur7DQ4 waveform \citep{2019PhRvR...1c3015V}; in
this case, the merger rate density is $1.7 \times
10^{-2}\,\text{Gpc}^{-3}\,\text{yr}^{-1}$. We find that all
GW231123-like events in our models originate exclusively from Pop III
binary stars. Furthermore, our calculated rate for GW190521-like
events matches the single-event rate estimate for GW190521
\citep[$0.13_{-0.11}^{+0.30}\,{\rm Gpc}^{-3}\,{\rm
    yr}^{-1}$;][]{2020ApJ...900L..13A}. Both Pop III binaries and
metal-poor stars ($Z < 10^{-4}$) contribute to the formation of
GW190521-like events, consistent with the results of
\cite{2022ApJ...926...83T}. In contrast, no GW231123-like events are
formed in the L3$\sigma$ set, despite its ZAMS-remnant mass relation
being similar to that of the M2$\sigma$ set. As summarized in Table
\ref{tab:Parameter_sets}, while the M models can produce GW231123-like
events provided the $\cago$ rate is at least $2\sigma$ below the
standard value, none of the L models are capable of generating such
mergers.

The physical reason for the disparity between the M and L models is as
follows. In the M model, a star with $\mzms = 150\,\msun$ expands only
up to $\sim 10^2\,\rsun$ during its post-main-sequence
phase. Consequently, binary systems consisting of two $150\,\msun$
stars lose only a negligible amount of mass ($\sim 0.01\,\msun$) via
mass transfer when their initial orbital separations are between $100$
and $200\,\rsun$. This minimal mass transfer occurs because the
progenitors of GW231123-like events experience stable mass transfer
owing to the radiative envelopes of the donors, and have mass ratios
close to unity. As a result, they leave behind BHs with masses nearly
equal to their ZAMS masses ($\sim 150\,\msun$) while maintaining their
orbital separations due to the lack of significant mass loss. Such
binary BHs are sufficiently compact to merge within the Hubble
time. In contrast, stars in the L model with $\mzms = 150\,\msun$
expand significantly, reaching radii of $\sim 3 \times
10^3\,\rsun$. Binary stars in this model undergo drastic mass
reduction to $\sim 70\,\msun$ through intense binary interactions --
primarily CE evolution.\footnote{This result is derived by analogy
with Pop I and II stars, although CE evolution in Pop III stars
remains highly uncertain.} These systems evolve into binary BHs with
component masses of $\sim 70\,\msun$, which are too small to be
classified as GW231123-like events. If the orbital separations are
wide enough ($\gtrsim 10^3\,\rsun$) to avoid such interactions, the
stars retain their masses; however, because they do not experience
orbital shrinking through binary interactions, their separations
remain too large for a merger to occur within the Hubble time.

\begin{figure}[ht!]
  \plotone{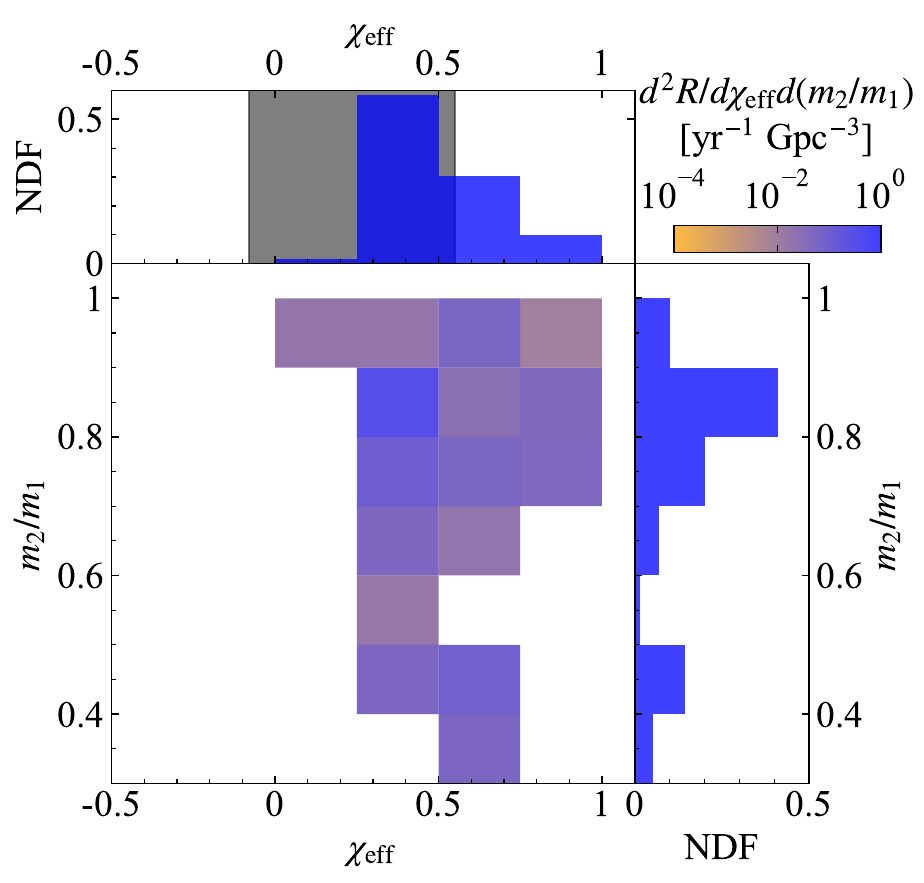}
  \caption{Distribution of the local merger rate density ($z=0$) for
    GW231123-like events as a function of the effective inspiral spin
    ($\xeff$) and mass ratio ($q = m_2/m_1$) in the M$2\sigma$
    set. The shaded region indicates the 90 \% credible interval for
    the effective inspiral spin, as derived from the combined model of
    the five pipelines used in \cite{2025arXiv250708219T}. The color
    bar of the $m_2/m_1$-$\xeff$ map is shown in the upper right. NDF
    stands for normalized distribution function.
    \label{fig:gw231123_xeff}}
\end{figure}

We pick up binary BHs which possess $\xeff$ and component masses
($m_1, m_2$) that are all consistent with the observed properties of
GW231123 (see the shaded region in the top-left panel of Figure
\ref{fig:gw231123_xeff}). From our initial population of 3 million Pop
III binaries, we identify 155 such events. Assuming a Poisson
distribution, the statistical uncertainty (standard error) for this
count is approximately 12. These events correspond to a merger rate
density of $1.3 \times 10^{-2}\,{\rm Gpc}^{-3}\,{\rm yr}^{-1}$. To
verify the statistical robustness of these results, we performed two
additional runs, each with 3 million Pop III binaries, using different
random seeds for the initial conditions. We find that the number of
corresponding binary BHs in these runs is 151 and 155, yielding merger
rate densities of $1.3 \times 10^{-2}$ and $1.5 \times 10^{-2}\,{\rm
  Gpc}^{-3}\,{\rm yr}^{-1}$, respectively.\footnote{Note that the
merger rate density is not strictly proportional to the number of
simulated binary BHs. The contribution of each binary depends on its
specific delay time, as the Pop III star formation rate is
time-dependent.}  Furthermore, the $\xeff$ and mass ratio
distributions remain consistent across all three runs. Based on these
findings, we conclude that our estimates for the merger rate density,
as well as the predicted $\xeff$ and mass ratio distributions for
GW231123-like events, are statistically robust.

To precisely determine the detailed parameter distributions of
GW231123-like events -- specifically those with effective inspiral
spins consistent with observations -- it would be necessary to evolve
significantly more than 3 million binaries or to implement an
efficient sampling technique, such as the STROOPWAFEL algorithm
\citep{2019MNRAS.490.5228B}. However, the primary objective of this
study is to demonstrate that a significant fraction of GW231123-like
events can indeed possess effective inspiral spins compatible with the
observed values. For this specific purpose, a yield of approximately
100 such events per 3 million simulated binary systems is sufficient
to confirm their formation as a viable evolutionary outcome.

In Figure \ref{fig:gw231123_xeff}, we investigate the distribution of
effective inspiral spins ($\xeff$) and mass ratios ($q = m_2/m_1$) for
the GW231123-like events in the M2$\sigma$ set.  In most of these
GW231123-like events, the primary and secondary BHs possess
dimensionless spins of $a_1 \sim 0$ and $a_2 \sim 1$,
respectively. This disparity arises from the following mechanism: the
progenitors of the primary BHs reach larger maximum radii compared to
those of the secondary BHs. Consequently, the primary BH progenitors
lose more mass than the secondary BH progenitors through stable mass
transfer. Although the total mass lost remains small -- typically less
than a few percent of the total stellar mass -- it is sufficient for
the primary BH progenitors to shed more spin angular momentum than the
secondary BH progenitors. This process occurs regardless of the
initial mass of the secondary progenitor; as a result, the mass ratios
of GW231123-like events span a wide range from $q \sim 0.3$ to $\sim
1$. Ultimately, our models yield $\xeff$ and $a_2$ values that are
consistent with the observational estimates for GW231123, even though
the predicted near-zero spins for the primary BHs ($a_1 \sim 0$) may
not fully align with some observational interpretations.

The GW231123-like events in our model exhibit semi-major axes of $\sim
100\,\rsun$ and orbital eccentricities of $\lesssim 0.5$ at the onset
of the binary BH stage, or at the moment when their binary evolution
finishes except for the orbital decay via GW radiation.  These
binaries will undergo significant orbital circularization before their
GW signals enter the millihertz frequency band. Consequently, these
sources could be observed as circular GW events by future space-borne
GW observatories, such as LISA \citep{2023LRR....26....2A}, TianQin
\citep{2016CQGra..33c5010L}, and Taiji \citep{2020IJMPA..3550075R}.

\section{Conclusion and Discussion}
\label{sec:Conclusion_and_Discussion}

We have examined the parameter space required to form GW231123-like
events, alongside other significant GW events. In this study,
GW231123-like events are defined as binary BH mergers whose component
masses fall within the 90 \% credible intervals of GW231123, as
derived from the combined model of five pipelines in
\cite{2025arXiv250708219T}. Our results demonstrate that GW231123-like
events can be formed from Pop III stars at a rate consistent with the
single-event rate estimate for GW231123, provided that convective
overshooting is inefficient and the $\cago$ reaction rate is at least
$2\sigma$ lower than the standard value. Furthermore, we find that
more than half of these simulated events possess effective inspiral
spins ($\xeff$) that are within the 90 \% credible interval for the
effective inspiral spin derived from the combined model of five
pipelines in \cite{2025arXiv250708219T}.

Indeed, our simulated GW231123-like events exhibit negligible
effective precessing spins ($\xpre \sim 0$). Within our chosen
parameter sets, there are no physical processes included that would
significantly tilt the BH spins; for instance, BHs receive only
minimal natal kicks during core collapse when mass loss is
limited. While this absence of precession initially appears
inconsistent with the observational data for GW231123, it is important
to note that determining effective precessing spins requires
higher-quality data than is necessary for constraining mass ratios or
effective inspiral spins. As \cite{2025arXiv250708219T} point out, the
current data do not allow for a confident claim of
precession. Furthermore, if we tentatively assume that the true
effective precessing spin of GW231123 is near zero, the observed
effective inspiral spin ($0.31_{-0.39}^{+0.24}$) being significantly
smaller than unity suggests that at least one of the component BHs may
not possess a high spin. In this context, our model's prediction --
where the primary BH has near-zero spin while the secondary is rapidly
rotating -- remains physically consistent with the overall properties
of GW231123.

The SFRD for Pop III stars adopted in this study are based on several
simulation results, while there are many other Pop III SFRD models
\citep[see review by][]{2023ARA&A..61...65K}. These integrated mass
densities are spread out between two digits, and our integrated mass
density is close to the median value of these models
\citep{2023MNRAS.524..307S}. Since the merger rate density of
GW231123-like events is proportional to the integrated mass density of
Pop III stars, our derived GW231123-like event rates may contain an
uncertainty of about two orders of magnitude.

The IMF for Pop III stars adopted in this study are based on several
simulation results, while the properties of Pop III binaries are
constructed by analogy with those of Pop I and II stars. These initial
conditions remain highly uncertain, given that Pop III stars have yet
to be observationally discovered. We find no GW231123-like events if
we assume the minimum period of Pop III binary stars is $10^{1.5}$
days, not $10^{0.15}$ days like the findings of
\cite{2012Sci...337..444S}. Similar results have been obtained by
\cite{2022ApJ...926...83T}. We set the minimum period to ten
  times that of our fiducial case because it is a convenient, round
  number and because an initial period of typically 10 days or less is
  representative of the progenitors producing GW231123-like events,
  thereby ensuring a clear contrast with our fiducial case. The
reason for no GW231123-like events is the lack of initially close
  binaries that can produce the binary BH systems compact enough to
  merge within a Hubble time as GW231123-like events.
This initial conditions may be justified by recent numerical
simulations \citep{2020ApJ...892L..14S, 2021MNRAS.508.4175C,
  2021MNRAS.501..643L}, which have shown that Pop III stars tend to
form wider binaries than Pop I/II stars. Nevertheless, note that these
simulations might not have enough resolution to form close Pop III
binary stars.

This uncertainty extends to both single and binary star evolution
models, as they cannot be calibrated against direct
observations. \cite{2023MNRAS.524..426I} have shown that binary
interaction parameters have great impacts on binary compact object
formation in the Pop I/II cases. Nevertheless, binary interaction
parameters do not affect the merger rate density of GW231123-like
events in our case. For the M2$\sigma$ set, we change a common
envelope parameter $\alpha$ from $1$ to $0.3$ and $3$, and the mass
transfer efficiency in the sub-Eddington regime from $0.5$ to $0.25$
and $1$. The GW231123-like (GW190521-like) event rates are $2.1$ --
$2.4\times 10^{-2}$ ($1.0^{-1}$ -- $1.5\times 10^{-1}$) ${\rm
  Gpc}^{-3}\,{\rm yr}^{-1}$, which are different from our fiducial
binary interaction parameters by only several 10 \%. While the ratio
of GW231123-like (GW190521-like) events to all the BH merger events is
$0.93 \times 10^{-3}$ ($4.4 \times 10^{-3}$) in our fiducial binary
interaction parameters, the ratios are $0.65$ -- $1.3 \times 10^{-3}$
($3.2$ -- $9.2 \times 10^{-3}$) in the other binary interaction
parameters. Their difference is at most factors of two. The reason for
the small difference is that GW231123-like events experience little
binary interaction.

Furthermore, the rapid population synthesis code employed here
utilizes a simplified treatment of stellar evolution. For instance,
stars that undergo mass gain or loss are assumed to evolve identically
to those of the same mass that have not experienced such
interactions. This approach may oversimplify the evolution of
interacting stars \citep{1983Ap&SS..96...37H, 1984Ap&SS.104...83H,
  2021ApJ...923..277R, 2025arXiv251220054X}.  Additionally, our model
does not account for the detailed evolution of internal stellar
structures during mass transfer; consequently, hydrogen envelopes are
assumed to be completely stripped during common envelope evolution,
which may be unphysical \citep{2021A&A...645A..54K}.  Nevertheless,
these simplifications likely have a minimal impact on our primary
results. In the M model, which is central to our findings, stellar
masses change by only a few percent throughout their evolution. These
stars experience negligible mass transfer due to their limited maximum
radii, and they do not trigger LBV-like winds. Ultimately,
establishing a plausible evolutionary scenario for Pop III binaries
remains a meaningful step toward demonstrating their potential as
progenitors of GW231123-like events.

Our results indicate that the progenitors of GW231123-like events
possess semi-major axes of $\sim 100\,\rsun$ and orbital
eccentricities of $\lesssim 0.5$ at the onset of the binary BH
stage. These orbits are expected to be fully circularized before their
GW signals enter the millihertz frequency band. Consequently, if such
events are detected by space-borne observatories like LISA, TianQin,
or Taiji, they should also be detectable as circular sources within
the LIGO, Virgo, and KAGRA bands \citep{2023PhRvX..13d1039A}. In
contrast, the dynamical formation channel in Pop III star clusters can
produce highly eccentric events with similar mass combinations, as
discussed in the companion paper \citep{2025ApJ...993L..30L}. Such
highly eccentric mergers might remain undetectable by millihertz
observatories even if they are prominent in the decahertz to kilohertz
bands. Therefore, the presence or absence of orbital eccentricity in
GW231123-like events provides a critical observational signature for
discriminating between the isolated binary and dynamical formation
origins of these massive mergers.

The discovery of GW231123 significantly narrows the plausible
parameter space for forming the observed population of binary BH
mergers through isolated binary evolution. Our analysis reveals that
GW231123 can originate from Pop III binaries only under the specific
conditions of inefficient convective overshooting and a $\cago$
reaction rate at least $2\sigma$ lower than the standard value. While
previous studies have shown that GW190521 can be produced from Pop II
binaries if the reaction rate is $3\sigma$ lower
\citep{2020ApJ...905L..15B}, or from Pop III binaries even with the
standard rate \citep{2021MNRAS.501L..49K, 2021MNRAS.505.2170T},
GW231123 imposes much stricter requirements. Ultimately, under the
assumption that the observed GW events are formed via isolated binary
evolution, these GW observations are now placing robust and
unprecedented constraints on the fundamental parameters of single-star
evolution models.

\begin{acknowledgments}
  We thank the anonymous referee for many fruitful suggestions. This
  research is supported partly by Grants-in-Aid for Scientific
  Research, 17H06360, 24K07040 and 25K01035 (A.T.). A.T. also thanks
  the Step-up program at Fukui Prefectural University. S.L.  thanks
  the support from Natural Science Foundation of China (Grant
  No. 12503054), Zhaoqing City Science and Technology Innovation
  Guidance Project (No. 250806120818306, No. 241216104168995) and the
  Young Faculty Research Funding Project of Zhaoqing University
  (No. qn202518). M.F. is supported by The University of Tokyo
  Excellent Young Researcher Program.  L.W. thanks the support from
  the National Natural Science Foundation of China through grant
  12233013, the Fundamental Research Funds for the Central
  Universities, Sun Yat-sen University (2025QNPY04), the High-level
  Youth Talent Project (Provincial Financial Allocation) through the
  grant 2023HYSPT0706.
\end{acknowledgments}

\begin{contribution}
  A. Tanikawa performed the calculations, analyzed the data, produced
  the figures, and led the manuscript writing. All authors contributed
  to the scientific discussion and manuscript writing.
\end{contribution}

\appendix

\section{Initial conditions of Pop I and II stars}
\label{sec:popIandII}

We simulated the evolution of Pop I and II binary stars in addition to
Pop III binaries to evaluate whether these diverse stellar populations
can reproduce the primary BH mass distributions inferred from GW
observations (see Figure \ref{fig:gw231123_rate}). In this Appendix,
we summarize the initial conditions adopted for Pop I and II
stars. For their SFRDs, we utilize the prescription from
\cite{2017ApJ...840...39M}, which is empirically derived from
rest-frame measurements in the X-ray, ultraviolet, and infrared bands.

Primary stars follow the Kroupa's IMF \citep{2001MNRAS.322..231K}
within the range of $0.08$--$150\,\msun$ for solar-metallicity stars,
while extremely metal-poor stars follow a top-heavy IMF with a
power-law index of $-1$ between $10$ and $150\,\msun$. For
intermediate metallicities, the IMF is modeled as a transition between
these two distributions, where the weight of the top-heavy component
increases as metallicity decreases, following the simulation results
of \cite{2021MNRAS.508.4175C}. Specifically, we adopt the transitional
IMF model described in \cite{2022ApJ...926...83T}: the Kroupa's IMF is
applied to 100 \%, 40 \%, 10 \%, and 0 \% of the primary stars for $Z
> 0.0002$, $2\times 10^{-4} < Z \le 0.0002$, $2\times 10^{-6} < Z \le
2\times 10^{-4}$, and $Z \le 2\times 10^{-6}$, respectively. The
remaining fraction of stars in each metallicity range follows the
top-heavy IMF. The binary fraction is assumed to be $0.5$, which is
broadly consistent with the findings of
\cite{2012Sci...337..444S}. For the distributions of binary mass
ratios, orbital periods, and eccentricities, we adopt the
observational constraints derived by \cite{2012Sci...337..444S}. To
implement these initial conditions, we consider a grid of
metallicities: $Z=0.02, 0.01, 0.005, 0.002, 0.001, 0.0005, 0.0002, 2
\times 10^{-6}$, and $2 \times 10^{-8}$. For each combination of IMF
and metallicity, we simulate $10^6$ binary systems to ensure
statistical convergence.

\section{Single star evolution model}
\label{sec:single_star}

Our stellar evolution model for $Z < 10^{-4}$ is constructed as
fitting formulae based on 1D hydrodynamic simulations performed with
the HOSHI code \citep{2016MNRAS.456.1320T, 2018ApJ...857..111T,
  Takahashi19, Yoshida19}. The HOSHI code incorporates convection,
semiconvection, and convective overshooting as chemical mixing
processes. The code includes a nuclear reaction network of 49 species
\citep{2018ApJ...857..111T}, a comprehensive stellar equation of state
\citep{Vardya60, Iben63, Blinnikov96}, OPAL, molecular and conductive
opacities \citep[][respectively]{Iglesias96, Ferguson05, Cassisi07},
and neutrino energy loss \citep{Itoh96}. Convective overshooting is
treated as a diffusive process extending beyond the convective
boundaries. The diffusion coefficient decreases exponentially with the
distance from the boundary, expressed as:
\begin{equation}
  D_{\rm cv}^{\rm ov} = D_{\rm cv,0} \exp [-2\Delta r/(f_{\rm
      ov}H_{\rm P0})],
\end{equation}
where $D_{\rm cv,0}$ and $H_{\rm P0}$ are the diffusion coefficient
and the pressure scale height at the convective boundary,
respectively, and $\Delta r$ is the distance from the boundary. The
overshooting parameter, $f_{\rm ov}$, is set to $0.01$ for the M$_{\rm
  A}$ model and $0.03$ for the L$_{\rm A}$ model, following the naming
convention in \cite{Yoshida19}. \cite{Yoshida19} demonstrated that the
evolution of a $20\,\msun$ star in the M$_{\rm A}$ and L$_{\rm A}$
models corresponds to the results obtained with GENEC
\citep{Ekstroem12} and Stern \citep{2011A&A...530A.115B},
respectively. Since stars in both cases -- particularly in the M$_{\rm
  A}$ model -- evolve into red supergiants, semiconvection is
moderately included in both the M$_{\rm A}$ and L$_{\rm A}$ models. As
for semiconvection in our models, it should be noted that we do not
consider highly efficient semiconvection. If semiconvection were
highly efficient, $20\,\msun$ stars would terminate their evolution as
blue supergiants without ever expanding to the red supergiant phase
\citep{1991A&A...243..155L, 2019A&A...625A.132S}.

As previously described, the HOSHI code can reproduce the results of
the GENEC and Stern models by adjusting the convective overshooting
parameters, corresponding to the M$_{\rm A}$ and L$_{\rm A}$ models,
respectively. This flexibility motivates the use of the BSEEMP code to
implement two distinct sets of fitting formulae for single star
evolution based on these models. For simplicity, we refer to these as
the M and L models, omitting the subscript ``A''. The fundamental
difference between the two lies in the efficiency of convective
overshooting: it is inefficient in the M model and efficient in the L
model. We find that models with less efficient overshooting (the M
model) result in smaller helium core masses at the end of the main
sequence and lower luminosities during the post-main-sequence
phase. Consequently, the maximum radius reached by the star is
significantly reduced. This effect is particularly pronounced in Pop
III stars; for instance, the maximum radius of an $80\,M_{\odot}$ star
is $\lesssim 10^2\,R_{\odot}$ in the M model, whereas it reaches
$\gtrsim 10^3\,R_{\odot}$ in the L model.

Our selection of inefficient convective overshooting (the M model) is
motivated by extensive studies utilizing similar assumptions, most
notably the GENEC models \citep{Ekstroem12, 2012A&A...542A..29G,
  2013A&A...558A.103G, 2019A&A...627A..24G, 2021MNRAS.501.2745M,
  2021MNRAS.502L..40F, 2021A&A...652A.137E, 2022MNRAS.511.2814Y,
  2024A&A...690A..91S}. While some observations of massive
main-sequence stars suggest that convective overshooting might be
efficient \citep{2014A&A...570L..13C, 2019A&A...622A..50H,
  2019A&A...625A.132S}, it remains highly valuable to investigate the
impact of lower overshooting efficiency on the formation and merger
rates of binary BHs. For comparison, our L model represents the case
of efficient convective overshooting and shares characteristics with
the PARSEC models \citep{2012MNRAS.427..127B, 2023MNRAS.525.2891C} in
addition to the Stern model \citep{2011A&A...530A.115B} mentioned
previously.

Massive stars with low metallicities can maintain high rotational
velocities due to reduced mass loss from stellar winds, which in turn
enhances rotational mixing. Such mixing might effectively mimic the
impact of convective overshooting by increasing helium core
masses. However, \cite{2021MNRAS.501.2745M} have demonstrated that for
Pop III stars with masses $\gtrsim 20\,\msun$, rotational mixing has a
negligible impact on key evolutionary properties -- including stellar
radii, total masses, and helium core masses -- provided that
convective overshooting is inefficient, as in our M model. While
rotational mixing can influence the evolution of lower-mass Pop III
stars ($\lesssim 20\,\msun$), particularly their radii
\citep{2021MNRAS.501.2745M}, this does not affect our findings. In our
model, these lower-mass stars ultimately form neutron stars rather
than BHs that are the focus of our study.

\section{Treatment of spin evolution}
\label{sec:spin_evolution}

We describe the treatment of stellar spin evolution, the details of
which are also provided in \cite{2022ApJ...926...83T}. The initial
stellar spin is assumed following the prescription of
\cite{2000MNRAS.315..543H}, which is calibrated to the main-sequence
rotation data from \cite{1992adps.book.....L}. Stellar spin evolves
through various processes, including stellar wind mass loss, wind
accretion, magnetic braking, stable mass transfer, and tidal
interactions. We adopt the evolution equations, moments of inertia,
and fitting formulae for gyration radii as described in
\cite{2002MNRAS.329..897H}. The key difference in our model lies in
the treatment of the radiative damping of the dynamical tide. While
\cite{2002MNRAS.329..897H} utilizes the formula from
\cite{1975A&A....41..329Z}, we adopt the updated formulations from
\cite{2010ApJ...725..940Y} and \cite{2018A&A...616A..28Q}, as
presented in Eqs. (20) and (21) of \cite{2022ApJ...926...83T}. In
these formulae, we assume that hydrogen-rich stars maintain a constant
convective core radius equal to their ZAMS value. This updated
treatment of radiative damping generally results in lower spins for
Pop III BHs \citep{2020MNRAS.498.3946K, 2022ApJ...926...83T}.

The spin angular momenta of the progenitors of GW231123-like events
are primarily governed by stable mass transfer and tidal interactions
rather than stellar wind mass loss, wind accretion, or magnetic
braking. These other processes have a negligible impact because the
progenitors experience extremely weak stellar winds, and minimal
magnetic braking, owing to their extremely low metallicity and small
maximum radii. While stable mass transfer is limited, it occasionally
impacts the spin angular momentum of the progenitors of GW231123-like
events. When the primary BH progenitors fill their Roche lobes, they
undergo slight mass loss. Since this mass is stripped from the
outermost layers, it results in a corresponding loss of spin angular
momentum. Although tidal interactions do spin up the progenitors, they
do not reach critical rotation rates for the following three
reasons. First, stars in the M model maintain small maximum radii
throughout their evolution. Second, we adopt the inefficient radiative
damping mechanism described previously, which limits the efficiency of
tidal angular momentum transfer. Third, since our focus is on binary
BH mergers in the local universe, these systems originate from
progenitors with relatively wide initial separations. Consequently,
the tidal spin-up experienced by these stars is comparatively weak
\citep{2016MNRAS.462..844K, 2017ApJ...842..111H}.

The spin angular momentum of a star is typically reduced during its
collapse to a BH, depending on the extent of the associated mass
loss. In our model, we assume that spin angular momentum is conserved
if the collapse proceeds without mass loss, following the formulation
in Eq. (15) of \cite{2022ApJ...926...83T}. If a calculated BH spin
parameter exceeds the physical limit of unity, we cap the value at
$1$. For the GW231123-like events identified in our simulations, the
progenitor stars do not undergo significant mass loss during the
collapse phase. Consequently, their BH spin angular momenta are
directly equivalent to those of their progenitors. It should also be
noted that these BHs do not experience substantial spin-up via stable
mass transfer, as the progenitors of GW231123-like events undergo very
little mass transfer, as previously discussed.

%\bibliography{natbib}{}
%\bibliographystyle{aasjournalv7}

\end{CJK*}
\end{document}